\documentstyle[aps,floats,graphicx,a4]{revtex}

\begin{document}

\title{Remarks on NonHamiltonian Statistical Mechanics:     \\
Lyapunov Exponents and Phase-Space Dimensionality Loss   }

\author{Wm. G. Hoover}
\address{Department of Applied Science       \\
  University of California at Davis/Livermore\\
  and Lawrence Livermore National Laboratory \\
  Livermore, California 94551-7808}

\author{H. A. Posch}
\address{Institute for Experimental Physics  \\
  University of Vienna, Boltzmanngasse 5     \\
  Vienna A-1090, Austria                      }

\author{K. Aoki}
\address{Department of Physics               \\
  Keio University                            \\
  4-1-1 Hiyoshi, Kohoku-ku                   \\
  Yokohama 223-8521, Japan                    }

\author{D. Kusnezov}
\address{Center for Theoretical Physics      \\
  Sloane Physics Laboratory                  \\
  Yale University                            \\
  New Haven, Connecticut 06520-8120           }

\date{\today}

\maketitle

\begin{abstract}

	The dissipation associated with nonequilibrium flow processes is
reflected by the formation of strange attractor distributions in phase
space.  The information dimension of these attractors is less than that
of the equilibrium phase space, corresponding to the extreme rarity of
nonequilibrium states. Here we take advantage of a simple model for heat
conduction to demonstrate that the nonequilibrium  dimensionality loss
can definitely exceed the number of phase-space dimensions required to
thermostat an otherwise Hamiltonian system.
\end{abstract}

\begin{center}
\end{center}
\vspace{0.5cm}
	Nonequilibrium molecular dynamics has been used to establish a
close link between microscopic dynamical phase-space instabilities and the
macroscopic irreversible dissipation associated with the Second Law of
Thermodynamics\cite{b1,b2}.  Both nonHamiltonian and Hamiltonian
methods have been used \cite{b2.5}.
The simplest such connection between
microscopic dynamics and macroscopic dissipation results when one or more
Nos\'e-Hoover thermostats\cite{b3} are used to control  nonequilibrium steady
states.  In the absence of nonequilibrium fluxes and with sufficient
phase-space mixing, these thermostats generate Gibbs' and Boltzmann's
canonical distribution.  In cases which include nonequilibrium driving
the instantaneous external entropy production rate
(due to heat transfer with the thermostats) is proportional to the sum of the
instantaneous Lyapunov exponents\cite{b4}:
$$\dot S/k \equiv -\sum \lambda = -\dot \otimes /\otimes \ .$$
The summed-up instantaneous exponents give the comoving rate of change
(following the motion) of an
infinitesimal phase-space hypervolume $\otimes $.  The sum of the
time-averaged Lyapunov exponents, $\{ \langle \lambda \rangle \} $
is necessarily negative, reflecting the irreversible dissipation
described by the Second Law of Thermodynamics.  

	Despite the time reversibility
of the underlying equations of motion, a symmetry breaking due to the
enhanced stability of trajectories proceeding ``forward'' in time relative
to their reversed twins, guarantees irreversibility and the formation of
a phase-space strange attractor\cite{b4}.  The dimensionality of such a
strange attractor can be measured by computing the smallest number of
time-averaged exponents (beginning with the largest one,
$\langle \lambda _1 \rangle $) with a negative sum.  This
``Kaplan-Yorke'' dimension is identical to the ``information dimension''
in cases of physical interest\cite{b5}.

	Because Liouville's Theorem\cite{b6} establishes that the
``extension in phase,'' the hypervolume of the comoving element $\otimes $,
cannot change in a motion governed by Hamiltonian dynamics, there has been some
reluctance to accept this strange-attractor explanation of irreversibility.
Very recently
Ramshaw has provided a particularly clear analysis of the generalized
Liouville's Theorem required to describe nonequilibrium dynamics with
time-reversible thermostats\cite{b7}. Refs. 9 and 10 provided a useful
computational
model and a means to estimate the dimensionality loss, $\Delta D$, the
difference between the strange attractor's information dimension and that
of the phase space in which it is embedded.  It was shown
that an accurate estimate for $\Delta D$ can be made, under certain
conditions, which requires only a single exponent, not the whole spectrum.
We use the whole
spectrum here because we want to make the demonstration of dimensionality loss
as convincing as possible.

	In the ``$\phi ^4$'' model an otherwise harmonic nearest-neighbor
lattice, with the Hooke's Law pair potential for Particles $i$ and $j$,
$$ \phi (r) = \textstyle{\frac{1}{2}}(r-1)^2\ ;
\ r = |r_i-r_j| > 0 \ ,$$
has each particle tethered to its lattice site with a quartic
potential, $\textstyle{\frac{1}{4}}\delta r^4$.  The quartic tethers 
prevent momentum
conservation, so that Fourier heat conduction can be observed, and also
can provide chaos, with one or more positive Lyapunov exponents.  To
model a nonequilibrium heat-conducting state two or more of the particles
are thermostated using feedback forces $\{ - \zeta p \} $ which are
linear or cubic in the
friction coefficient $\zeta $ and the momentum $p$.  Here we choose the
simplest case, linear in both variables, the Nos\'e-Hoover
thermostat.  In one space dimension the thermostated equations of motion
are:
$$\dot p = F - \zeta p\ ;\dot \zeta = [(p^2/mkT)-1]/\tau^2 \ ,$$
where $\tau $ is the characteristic response time of the thermostat
force, $-\zeta p$.  In two dimensions, where $p^2$ is a sum of $x$ and
$y$ components, $T$ is replaced by $2T$.
The target temperature $T$ for the thermostat is necessarily achieved
whenever a stationary solution exists, such that the long-time-averaged
value of $\dot \zeta $ vanishes.  Because the long time average of
$\zeta \dot \zeta = (d/dt)\frac{1}{2}\zeta^2 $ likewise vanishes,
the heat transferred by any
thermostated momentum can be directly related to the temperature:
$$\langle \zeta p^2/m\rangle \equiv  \langle \zeta \rangle kT \ ,$$
again explicitly writing only the one-dimensional case.

\begin{figure}
\centering{\includegraphics[width=8cm,angle=-90,clip=]{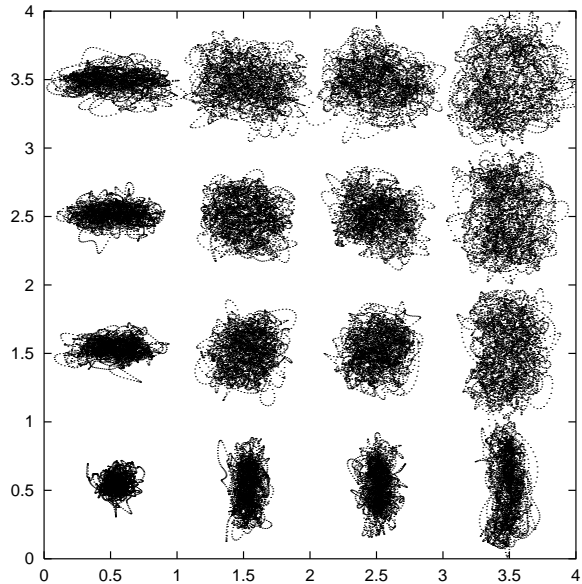}}
\caption{Stationary state particle trajectories for a 16-particle 
system with dimensionality
loss of 12.5 induced by two Nos\'e-Hoover thermostated particles at
temperatures of 0.001 and 0.009.}
\label{figure1}
\end{figure}

\begin{figure}
\centering{\includegraphics[width=8cm,clip=]{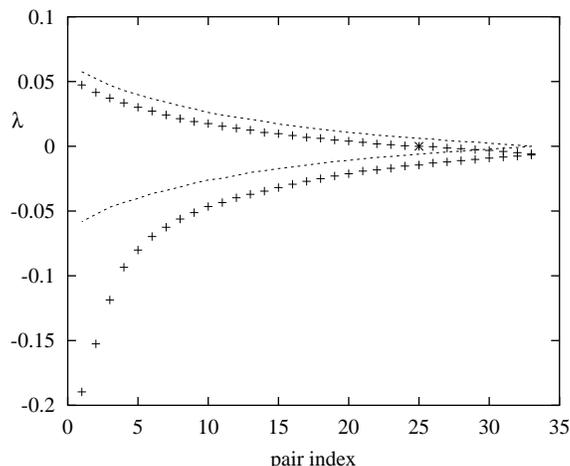}}
\caption{Lyapunov spectrum for the system shown in Fig. \ref{figure1}
There are 24 positive exponents, a single vanishing exponent (emphasized
here), and 41 negative exponents describing the motion in the 66-dimensional
space.  In the symmetric equilibrium situation (with the hot and cold
temperatures equal) the exponents occur in 33 pairs, $\{ \pm \lambda \} $,
with a single pair of vanishing exponents.  In the nonequilibrium case 
there is an overall shift toward more negative values,
$\dot S/k \equiv -\sum \lambda > 0 $.  }
\label{figure2}
\end{figure}

The 16-body system shown in Fig. \ref{figure1} has a cold particle at 
the lower left
corner, a hot particle at the upper right corner, along with 14 Newtonian
particles able to transmit heat from the hot particle to the cold one.
This system (as well as many others) can exhibit dimensionality losses
exceeding the 10 phase-space coordinates $\{ x,y,p_x,p_y,\zeta \} $
required to describe the two thermostated particles.  In the case shown
in the Figure, with cold temperature 0.001 and hot temperature 0.009, and
a Nos\'e-Hoover relaxation time $\tau = 1.4$,
the dimensionality loss is 12.5.  This means that the sum of the 53
largest Lyapunov exponents, plus half the 54th, is zero.  
See Fig. \ref{figure2}
for the spectrum of exponents.  The strange
attractor has an information dimension of 53.5, embedded in a 66
dimensional phase space, of which 56 dimensions represent the
purely Hamiltonian particles.

This result establishes very
clearly that the
dissipation and overall contraction occurring in the thermostated part of
the system can cause a loss in the Hamiltonian region too.  Although we
had been able to simulate systems with dimensionality losses barely
exceeding the additional boundary phase-space coordinates\cite{b10},
the present
results are much more clearcut.  Otherwise one might imagine that the
thermostat regions simply provide a time-dependent force, which for
Hamiltonian dynamics can provide no dimensionality loss.  The {\em feedback}
linking the forces to the phase-space coordinates, not just to the time,
makes
the loss possible.

	This situation is analogous to the dynamics of a one-dimensional
damped harmonic oscillator with a {\em fixed} friction coefficient $z$, for
which the logarithmic phase-space contraction rate,
$$\dot \otimes /\otimes \equiv (\partial \dot q/\partial q) +
  (\partial \dot p/\partial p)\ ,$$
occurs in the
momentum direction only:
$$\dot q = p\ ;\ \dot p = -q - zp \longrightarrow $$
$$(\partial \dot q/\partial q) = 0\ ;\
  (\partial \dot p/\partial p) = - z\ .$$
The damped oscillator motion nevertheless collapses
$(\otimes \rightarrow 0)$ to the fixed point at
the origin, $(q=0,p=0)$ because the motion {\em rotates} the comoving phase
volume.  Phase-space rotation\cite{b11} in many-body systems increases
rapidly with
system size and is necessarily the mechanism through which the volume
collapse can spread to directions without direct contraction.

	In summary, thermostated heat flow, in a tethered harmonic lattice,
shows conclusively that the dimensionality loss associated with 
irreversible processes can exceed the additional variables required to
specify the thermal boundaries driving the system from equilibrium.  This
shows that the dimensionality loss is a real feature of nonequilibrium
systems, not an artifact of a particular choice of thermostat.  We
believe that this finding clarifies a point which has been strenuously
debated over the past decade\cite{b12}.  There can be no doubt that
advances in parallel computing will soon elucidate the precise way in which
the large-system limit is obtained.

        The approach we follow here connects 
irreversible entropy production to multifractal phase-space structures
through time-reversible Nos\'e-Hoover thermostats. It is worth mentioning
that there are now alternative connections, including some based on purely 
Hamiltonian mechanics \cite{b2.5}.

	We thank Aurel Bulgac, Carol Hoover, Rainer Klages, and John Ramshaw
for constructive discussions.  WGH's work in Carol Hoover's
Methods Development Group at the Lawrence Livermore National Laboratory was
performed under the auspices of the United States Department of Energy
through University of California Contract W-7405-Eng-48.  HAP's work
was supported by the Austrian Fonds zur F\"orderung der wissenschaftlichen
Forschung, Project P-15348.


%

\end{document}